\documentstyle[prl,aps,epsf,twocolumn]{revtex}
\def\break#1{\pagebreak \vspace*{#1}}
\begin{document}
\draft
\title{Coulomb charging energy for arbitrary tunneling strength}
\author{Xiaohui Wang, Reinhold Egger, and Hermann Grabert}
\address{Fakult{\"a}t f{\"u}r Physik, 
Albert-Ludwigs-Universit{\"a}t Freiburg,
Hermann-Herder-Stra{\ss}e  3, D-79104  Freiburg}
\maketitle
\widetext
\begin{abstract}
The Coulomb energy of a small metallic island coupled
to an electrode by a tunnel junction is investigated.
We employ Monte Carlo simulations to determine
the effective charging energy for arbitrary tunneling strength. 
For small tunneling conductance, the data agree with analytical
results based on a perturbative treatment of electron tunneling,
while for very strong tunneling recent semiclassical results 
for large conductance are approached. The 
data allow for an identification of the range of 
validity of various analytical predictions.
\end{abstract}


\narrowtext
Charging effects in metallic nano\-structures \cite{gra92}
continue to be studied extensively both theoretically and 
experimentally. While for the case of weak tunneling 
the underlying physics is well understood \cite{ing92}
and starts to find metrological applications \cite{nie96},
for the case of strong tunneling there are several
conflicting analytical predictions
\cite{gui86,mat91,pan91,zw91,gra94a,gol94,gra94b,sch94,fal95,heins,wang96}.
To settle these discrepancies, we have performed precise Monte Carlo
calculations for a large range of the tunneling strength.

\begin{figure}
\epsfysize=9cm
\epsffile{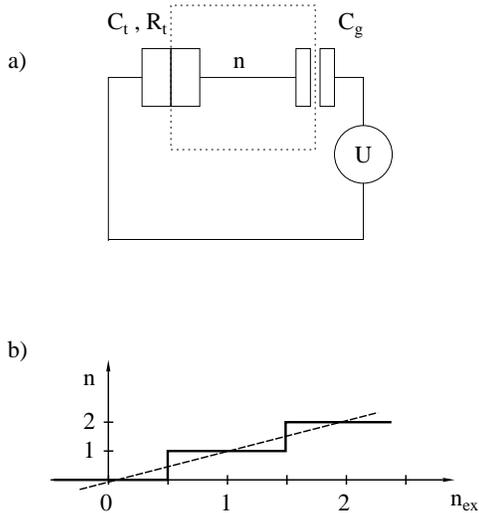}
\caption[]{\label{fig1} a) The circuit diagram of the
single-electron box, consisting of a tunnel junction in
series with a capacitor $C_g$. The junction resistance is
$R_t$ and the junction capacitance is $C_t$. b) 
The number $n$ of electrons in
the box as a function of the applied voltage $U=(e/C_g) n_{\rm ex}$ in the
Coulomb blockade regime, $R_t \gg R_K$ (solid line) and
for strong tunneling, $R_t \ll R_K$ (dashed line).}
\end{figure}

The simplest device displaying Coulomb charging effects 
is the single electron box (SEB) which is formed by a metallic island 
between a tunnel junction and a gate capacitor, see Fig.~\ref{fig1}.
The external electrodes are biased by a voltage source.  
If the tunneling resistance $R_t$ is large compared to
$R_K = h/e^2 \sim 25.8\ {\rm k}\Omega$, the charge $q$ on the
metallic island is quantized, $q=-ne$, where $n$ is the
number of excess electrons on the island. A simple electrostatic
calculation for vanishing electron tunneling gives for the
ground state an electron number $n$ which is the integer
closest to $n_{\rm ex} = C_g U/e$. Hence,
as a function of the applied voltage, $n$ displays the well-known
Coulomb staircase.

At finite temperature the Coulomb staircase is rounded by
thermal fluctuations of the island charge. This thermal
smearing is easily understood and seen experimentally \cite{laf91}.
However, there is also a quantum-mechanical rounding of the
steps, since the island charge is not truly quantized. Electron
tunneling leads to hybridization of the states in the lead and
island electrodes. For very strong tunneling, we 
expect a complete washout of charging effects, and the average
electron number $\langle n \rangle$ in the box becomes
proportional to the applied voltage,
$\langle n \rangle = n_{\rm ex}$. As indicated in Fig.~\ref{fig1}, there is a 
gradual breakdown of the Coulomb blockade behavior
as the dimensionless tunneling conductance $\alpha_t = R_K/R_t$ 
increases.
 
The classical single electron charging energy of the SEB is 
given by $E_c=e^2/2C$, where the island capacitance $C$ is the 
sum of the capacitance $C_t$ of the tunnel junction and the gate 
capacitance $C_g$ (see Fig.~\ref{fig1}). This quantity can be extracted
\break{1.3in}
from measurements in the classical regime \cite{laf91}.
The strength of the Coulomb blockade effect may be
described in terms of an effective charging energy
$E_c^*$ which coincides with $E_c$ in the limit of
small quantum fluctuations of the charge, $\alpha_t \ll 1$,
and vanishes in the limit of strong tunneling,
$\alpha_t \gg 1$. Since in the latter case 
$\partial \langle n \rangle /\partial n_{\rm ex} \to 1$,
we may define $E_c^*$ from the slope
of the staircase at $n_{\rm ex}=0$ by
\[
E_c^* = E_c^{}\left( 1- \partial \langle n \rangle / 
\partial n_{\rm ex} \vert_{n_{\rm ex}=0} \right) \;.
\]
A completely equivalent definition of $E_c^*$ in
terms of the free energy $F(n_{\rm ex},\alpha_t)$ of the SEB reads
\begin{equation}\label{eqn2}
E_c^* = \left. \frac12 \frac{\partial^2 F(n_{\rm ex},\alpha_t)}{
 \partial n_{\rm ex}^2} \right|_{n_{\rm ex}=0} \;.
\end{equation}

Several previous articles have made analytical predictions 
on $E_c^*/E_c^{}$ as a function of $\alpha_t$. For small
$\alpha_t$, perturbation theory in the tunneling term may 
be employed. One finds at zero temperature \cite{foot}
\[
 E_c^*/E_c = 1 - (1/\pi^2) \alpha_t + d_2 \alpha_t^2
 + {\cal O}(\alpha_t^3)\; .
\]
The linear term is readily evaluated \cite{mat91}. For 
higher-order corrections one must take into account
that the charging energy in the unperturbed 
Hamiltonian leads to a correlation of
the Fermi liquids in the lead and island electrodes. 
Based on the non-crossing approximation (NCA), 
Golubev and Zaikin \cite{gol94} find 
(see Eq.~(8) in Ref.~\cite{gol94})
\begin{eqnarray}
d_2 &=& {1\over 8\pi^4}\bigg({{7\pi^2-64}\over 12} 
+{{29\ln(2)}\over 9}  
- 4 \ln(2)\ln(3)
+ 2 \hbox{Li}_2(3/4)\bigg) \nonumber \\
\label{gol}
 &=& 3.134/16\pi^4 \;, 
\end{eqnarray}
where Li$_2(x) = -\int_0^x dz\, \ln(1-z)/z$ is the dilogarithm
function. On the other hand, from the systematic evaluation 
of all diagrams, Grabert \cite{gra94a,gra94b} obtains
(see Eq.~(2) in Ref.~\cite{gra94a} or Eq.~(100) in 
Ref.~\cite{gra94b})
\begin{eqnarray} \nonumber
d_2 &=& {1\over 8\pi^4} \bigg(  
{{4\pi^2-16}\over 3} + {{32 \ln (2)}\over 9} 
 - 8 \ln^2 (2) 
- 4 \hbox{Li}_2(3/4)  \bigg) \nonumber \\
\label{gra}
    &=& 5.066/16 \pi^4 \;.
\end{eqnarray}
The difference between Eqs.~(\ref{gol}) and (\ref{gra}) 
indicates that the NCA becomes a poor approximation for large $\alpha_t$. 

The case of strong tunneling, $\alpha_t \gg 1$,
was first studied by Panyukov and  Zaikin \cite{pan91}. 
Based on an instanton approach \cite{foot2},
these authors find (see Eqs.~(7) and (8) in Ref.~\cite{pan91})
\begin{equation} \label{pz}
E_c^*/E_c = \alpha_t^2 \exp(-\alpha_t/2 + \gamma) \;,
\end{equation}
where $\gamma =0.577\ldots$ is Euler's constant.
Very recently, Wang and Grabert
have evaluated the low-temperature partition function
of the SEB by related path-integral methods\cite{wang96}.
For large $\alpha_t$, the path integral is dominated by
multi-sluggon trajectories whose contribution
can be summed analytically yielding (Eq.~(10) in
Ref.\ \cite{wang96})
\begin{equation} \label{wg}
E_c^*/E_c=2  \alpha_t^3e^{-\alpha_t/2} 
[1+{\cal O}(\ln (\alpha_t)/\alpha_t)] \;.  
\end{equation}
The exponential dependence of $E_c^*$ on $\alpha_t$ 
in Eqs.~(\ref{pz}) and (\ref{wg}) is
in accordance  with a recent renormalization-group analysis 
\cite{fal95}. However, the
pre-exponential factors differ by orders of
magnitude in the strong-tunneling regime.
Available renormalization-group techniques do
not resolve this problem.

To clarify the discrepancies and identify 
reliable methods for further analytical work, 
we have performed Monte Carlo (MC) simulations. 
We start from the path integral for the partition
function of the SEB which may be written as
\begin{equation} \label{parts}
Z(n_{\rm ex}, \alpha_t) = \int D [\varphi] e^{-S_{\rm box}[\varphi]} \;, 
\end{equation}
where $\beta=1/k_BT$ is the inverse temperature, and
the  integral is over all paths of the phase
$\varphi(\tau)$ in the interval $-\beta/2 \leq \tau \leq \beta/2$ 
with $\varphi(\beta/2)=\varphi(-\beta/2)$ modulus  $2\pi$. The action 
\[
S_{\rm box}[\varphi]=S_c[\varphi]+S_t[\varphi]
\]
contains two parts describing charging of the island and 
tunneling across the junction.  The effect of the Coulomb energy is 
contained in 
\[
S_c[\varphi]=\int_{-\beta/2}^{\beta/2} d\tau \left[ \frac
{1}{4E_c}(\dot{\varphi}+2in_{\rm ex}E_c)^2+E_cn_{\rm ex}^2 \right] \;.
\]
The second part of the action\cite{ben83}
\[
S_t[\varphi]=2\int_{-\beta/2}^{\beta/2} 
d\tau \int_{-\beta/2}^{\beta/2} d\tau' \alpha(\tau-\tau') 
\sin^2 \left[ 
\frac{\varphi(\tau)-\varphi(\tau')}{2} \right]
\]
describes tunneling. The  Fourier transform of $\alpha(\tau)$ reads
$\alpha_l=-\alpha_t |\omega_l|/4\pi$ for $|\omega_l| \ll D$,
where $\omega_l=2\pi l/\beta$ are the Matsubara frequencies and $D$ is 
the electronic bandwidth.

By using the definition (\ref{eqn2}) for the effective charging 
energy, we can relate $E_c^*$ to the partition function,
\begin{equation}\label{ecstern}
  E_c^* = \left.  -\frac{1}{2\beta} \frac{\partial^2 
\ln Z(n_{\rm ex},\alpha_t)}
{\partial n_{\rm ex}^2} \right|_{n_{\rm ex}=0}\;.
\end{equation}
To proceed, we write Eq.~(\ref{parts}) in the equivalent form
\begin{equation} \label{partsn}
Z(n_{\rm ex}, \alpha_t) = \sum_{m=-\infty}^\infty
e^{2\pi i m n_{\rm ex}} 
\int D[\vartheta] \exp(-S_m[\vartheta])\;,
\end{equation}
where we have separated the functional integration into a 
sum over winding numbers $m$ and an integration over all
paths $\vartheta(\tau)$ which are now subject to the boundary conditions
$\vartheta(-\beta/2)=\vartheta(\beta/2)=0$. Here,
the action $S_m[\vartheta]$ is given by
\begin{eqnarray}\label{sm}
&& S_m[\varphi]  = \frac{\pi^2 m^2}{\beta E_c} + \frac{1}{4 E_c}
\int d\tau \, \dot{\vartheta}^2(\tau)\\ & +&2 \int d\tau \int d\tau^\prime
\alpha(\tau-\tau^\prime) \sin^2 \left[ 
\frac{\vartheta(\tau)-\vartheta(\tau')}{2}+
 \frac{\pi m (\tau-\tau^\prime)}{\beta} \right] \;. \nonumber
\end{eqnarray}
>From Eqs.~(\ref{ecstern}) and (\ref{partsn}),
a convenient starting  point for MC simulation is given by
\begin{equation}\label{mcs}
\frac{E_c^*}{E_c}  = \frac{2\pi^2}{\beta E_c}\; \frac{
\sum_m m^2 \int D[\vartheta] \exp(-S_m[\vartheta]) }
{\sum_m \int D[\vartheta] \exp(-S_m[\vartheta]) }\;.
\end{equation}
which expresses $E_c^*$ in terms of the mean squared 
winding number.

\begin{figure}
\epsfysize=8cm
\epsffile{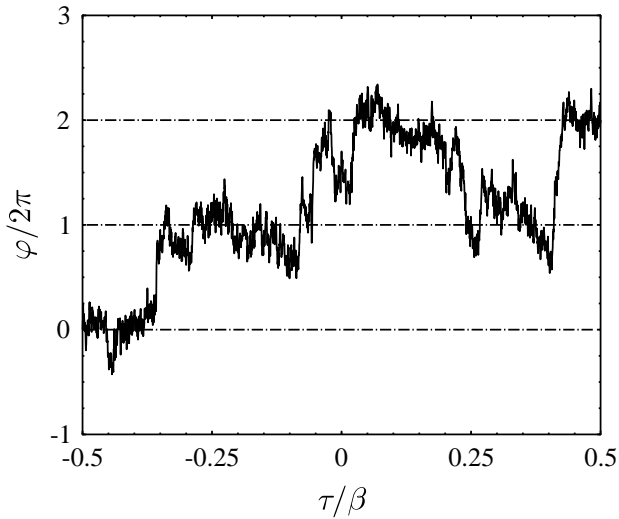}
\caption[]{\label{fig2} Typical paths $\varphi(\tau)$ generated
in the MC simulation for large $\alpha_t$ and low temperatures. Specifically,
this configuration was present after 50,000 MC steps for
$\alpha_t=30$ and $\beta E_c=500$, taking $\varphi(\tau)=0$
initially.}
\end{figure}

For the MC simulation, we discretize imaginary time in
$P$ slices $\tau_i$  of sufficiently small length $\Delta \tau=\beta/P$,
and then sample $\vartheta(\tau_i)$ 
and the winding number $m$ from the action $S_m[\vartheta]$ specified
in Eq.~(\ref{sm}). Measuring $m^2$ along the MC trajectory  
allows to extract $E_c^*$ according to Eq.~(\ref{mcs}).
As is apparent from Eq.~(\ref{partsn}), the
MC weight function is not necessarily positive definite 
for non-integer $n_{\rm ex}$. The resulting interference can
lead to numerical instabilities,  
especially near half-integer values of $n_{\rm ex}$. 
The MC algorithm developed in Refs.~\cite{fal95,heins} deals with 
the case of general $n_{\rm ex}$ and hence suffers from this problem.
However, since we focus on the effective charging energy 
only and hence on the value $n_{\rm ex}=0$,
there is no instability problem in our algorithm. 
This circumstance allowed us to reach temperatures of about one
order of magnitude lower than studied in Ref.~\cite{fal95},
typically $\beta E_c = 500$.

For a given parameter set $(\alpha_t, \beta E_c)$, we have first 
determined the Trotter number $P$ by empirically checking
convergence to the large-$P$ limit. Typically, a value of
$P=5 \beta E_c$ was sufficient. Special care is then
necessary for large $\alpha_t$ since the acceptance rates
can be very low. The 
most important paths $\varphi(\tau)$ encountered in this regime are of the 
form shown in Fig.~\ref{fig2}. These paths closely resemble
the multi-sluggon trajectories contributing to the semiclassical
result (\ref{wg}) \cite{wang96}. The predominant occurrence of these
paths in the MC sampling gives a strong indication that
the semiclassical sluggon calculus is indeed appropriate for
large $\alpha_t$.
The MC updating employs single-particle moves, where different samples are 
separated by five passes. Results reported below are obtained from
several million samples per parameter set. The simulations were carried out
on IBM RISC 6000/590 workstations.

\begin{figure}
\epsfysize=10cm
\epsffile{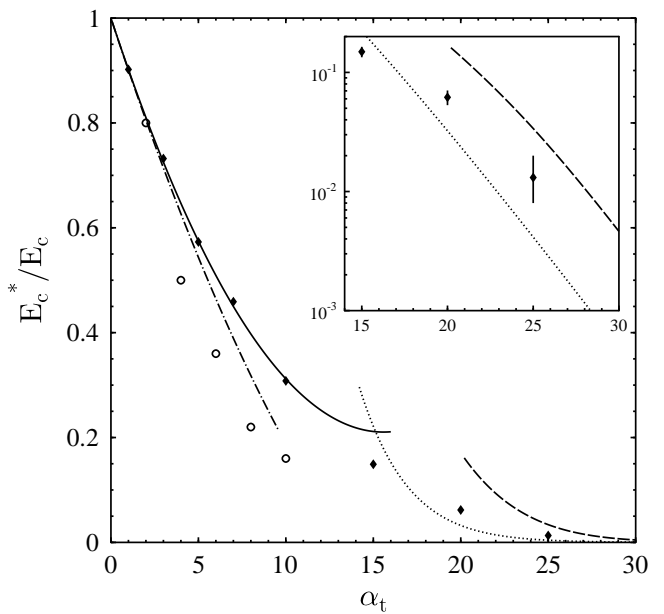}
\caption[]{\label{fig3}  MC data for $E_c^*$ as 
a function of $\alpha_t$ and comparison to previous work. The perturbative 
results for small $\alpha_t$ are depicted as the
dashed-dotted curve (see Golubev and Zaikin \cite{gol94}) and  the
solid curve (see Grabert \cite{gra94a,gra94b}), respectively.
The analytical results for large $\alpha_t$ are the
dotted curve (for zero temperature, see Panyukov and Zaikin \cite{pan91}) and
the dashed curve (for $\beta E_c=500$,
see Wang and Grabert \cite{wang96}), respectively.
The MC data by Falci {\em et al.}\cite{fal95} are shown as
open circles, and our MC data are given by filled diamonds.
In the inset, the data for large $\alpha_t$ together with the
curves from Refs.\cite{pan91,wang96} are replotted 
on a semi-logarithmic scale.
The data points were obtained at $\beta E_c=100$  for
$\alpha_t<15$ and at $\beta E_c=500$ for $\alpha_t\geq 15$. 
Statistical errors on our MC data are smaller than the symbol size unless
indicated by vertical bars.  }
\end{figure}

Our data for the effective charging energy are shown in Fig.~\ref{fig3}
together with the various analytical predictions 
\cite{pan91,gra94a,gol94,gra94b,wang96}
and the MC results by Falci {\em et al.}\cite{fal95}.  
For weak tunneling, 
the zero-temperature limit was approached already
for rather high temperatures, $\beta E_c=100$. However, 
it turns out to be quite 
a demanding task to reach this limit for large $\alpha_t$. 
In fact, the data points for $\alpha_t=20$ and 
$\alpha_t=25$ (which were obtained for $\beta E_c=500$)
are {\em not}\, in the true zero-temperature limit yet. This 
follows from comparing additional MC data  obtained at
different temperatures (not shown here) and has
to be taken into account when comparing the finite-temperature
MC data with the zero-temperature
analytical predictions (\ref{pz}) and (\ref{wg}).

For the case of {\em weak tunneling}, the theoretical predictions
by Grabert \cite{gra94a,gra94b} given in Eq.~(\ref{gra}) are
confirmed with very good precision up to surpringly large 
values of the tunneling conductance, $\alpha_t\leq 10$,
while the NCA result (\ref{gol}) becomes inaccurate for 
$\alpha_t > 5$.  We suspect that the discrepancy between
the MC data of Ref.~\cite{fal95} 
and our results is due to the lower temperatures
employed here. For $\alpha_t\leq 15$, our simulations have 
converged to the zero-temperature limit.

For the case of {\em strong tunneling}, the predictions 
of Panyukov and Zaikin \cite{pan91} and of the sluggon calculus \cite{wang96}
differ by about one order of magnitude due to different pre-expontial factors,
while our numerical data for $\beta E_c=500$
 lie in between both predictions \cite{foot3}. Since a 
further decrease of temperature will suppress thermal fluctuations
even more, our MC results give a lower bound to
the true zero-temperature result for $E_c^*$.
However,  the values of $\alpha_t$ are still not
large enough to confirm the sluggon prediction (\ref{wg}) 
unambiguously.

In conclusion, we have developed and carried out Monte Carlo simulations
for the effective charging energy of the single electron box. The 
data describe a gradual smooth crossover between the analytically
accessible limits of weak tunneling (where
 perturbation theory in the tunneling
conductance applies) and strong tunneling (where a semiclassical 
calculus is appropriate).

The authors would like to thank G.~Falci,
G.~Sch{\"o}n, and W.~Zwerger for fruitful discussions.
Financial support was provided by the
Deutsche Forschungs\-ge\-mein\-schaft.


\begin{references}
\bibitem{foot} We note that the dimensionless conductance
$\alpha_t$ is related to the parameter $g$ used in 
Refs.\ \cite{mat91,gra94a,gol94,gra94b} by $g= \alpha_t/4 \pi^2$.
\bibitem{foot2} The dimensionless conductance
$\alpha_t$ used by Panyukov and Zaikin (PZ) \cite{pan91} 
is related to the tunneling conductance used here by
$\alpha_t^{\rm PZ} = \alpha_t/4$.
\bibitem{foot3} As the error bars on the MC data by Falci 
{\em et al.}~\cite{fal95} are very large for strong tunneling,
we have only displayed their data for $\alpha_t \leq 10$ in
Fig.~\ref{fig3}.
\end{references}
\end{document}